# Equilibria of Plurality Voting with Abstentions

Yvo Desmedt*      Edith Elkind[†]

October 24, 2018

**Abstract**

In the traditional voting manipulation literature, it is assumed that a group of manipulators jointly misrepresent their preferences to get a certain candidate elected, while the remaining voters are truthful. In this paper, we depart from this assumption, and consider the setting where all voters are strategic. In this case, the election can be viewed as a game, and the election outcomes correspond to Nash equilibria of this game. We use this framework to analyze two variants of Plurality voting, namely, simultaneous voting, where all voters submit their ballots at the same time, and sequential voting, where the voters express their preferences one by one. For simultaneous voting, we characterize the preference profiles that admit a pure Nash equilibrium, but show that it is computationally hard to check if a given profile fits our criterion. For sequential voting, we provide a complete analysis of the setting with two candidates, and show that for three or more candidates the equilibria of sequential voting may behave in a counterintuitive manner.

## 1 Introduction

Voting and elections play an important part in the functioning of the human societies, and hold a lot of promise as a tool for preference aggregation in multiagent systems [7]. However, when voters are strategic, i.e., misreport their preferences in order to get their favorite alternative elected, the outcome of a voting procedure may not reflect the true preferences of the society as a whole. Unfortunately, the seminal work of Gibbard and Satterthwaite [9, 15] implies that any election system is prone to manipulation by the voters. For this reason, the study of manipulability of various voting rules, i.e., determining the fraction of manipulable profiles or understanding the algorithmic complexity of individual or coalitional manipulation, is an active research area (see, e.g., [6] for a sample of recent work). Much of this work views manipulation as a type of adversarial behavior that needs to be prevented, either by imposing restrictions on voter's preferences, or identifying a voting rule for which manipulation is computationally hard, preferably in the average case rather than in the worst case. Further, it is usually assumed that there is either a unique manipulator or a group of manipulators, which know how everyone else votes and select their vote(s) accordingly; the remaining voters are assumed to vote non-strategically.

We can, however, take the approach suggested in [17, 5, 2] (see also Section 6) and view manipulation from a different perspective, namely, as an unavoidable attribute of an electoral system

---

*Department of Computer Science, University College London, UK and Research Center for Information Security, AIST, Japan

[†]Division of Mathematical Sciences, School of Physical and Mathematical Sciences, Nanyang Technological University, Singapore

with rational voters. From this perspective, we do not classify the voters as honest or manipulative, but rather assume that all voters act strategically. Voting then becomes a game where players are voters and actions are votes, and the outcome of the election can be identified with a Nash equilibrium of this game. In this context, voting rules should be analyzed according to their behavior in equilibrium, rather than under assumption of truthful voting, as it is usually done. Further, as mixed equilibria are commonly deemed to be unacceptable in the context of voting, an important issue in this framework is to identify the preference profiles that possess pure Nash equilibria.

In this paper, we apply such analysis to two variants of Plurality voting (i.e., the most basic voting rule, where each voter votes for a single candidate), namely, simultaneous voting and sequential voting. As in the overwhelming majority of the literature on voting manipulation, we assume that the voters know each others' true preferences. However, they do not necessarily know each others' votes. Specifically, in the simultaneous voting model, where all voters submit their votes at the same time, no voter knows how others vote. On the other hand, in the sequential voting model the voters vote one by one in an exogenously determined order, and each voter can observe the actions of the voters who voted before him. We are interested, respectively, in Nash equilibria and subgame-perfect Nash equilibria (for definitions, see Section 2) of the resulting games. We also make an additional assumption that voting has a small cost and the voters can abstain in order to save this cost (interestingly, it turns out that in the sequential setting the voters may use this option for strategic reasons). This assumption has been previously used in the analysis of voting behavior by Battaglini [2], and allows us to eliminate some of the "undesirable" equilibria.

For the simultaneous voting model, we give a complete characterization of preference profiles that admit pure Nash equilibria. However, we show that checking whether a particular profile fits our criterion is computationally hard. On the other hand, the problem becomes polynomial-time solvable if the number of candidates or the number of voters is small.

We then turn our attention to the sequential voting model. Our interest in this model is motivated by numerous real-life scenarios where voters may observe others' votes before they need to cast their ballot, such as the roll call voting in the U.S. Senate or House of Representatives, or popular voting in large countries that span several time zones, where the voters in the west may already know the outcome in the eastern regions by the time they need to vote. Another example is provided by the U.S. primaries, where the voting is spread over several months. In such settings, the early voters may vote strategically to influence subsequent voters, and late voters take into account the votes already cast, so the outcome of such elections is likely to be quite different from that of their simultaneous counterparts. In this paper, we show that this is indeed the case. We first provide a complete analysis of sequential elections with two candidates, showing that their outcomes have many desirable properties (the voters are essentially truthful, the majority-preferred candidate always wins). However, we demonstrate that our analysis for the two-voter case does not generalize to three or more voters. Indeed, sequential voting with three or more voters exhibits several counterintuitive properties: a voter may vote for his least preferred candidate or strategically abstain, as well as vote for a candidate who ends up not winning. We also discuss the complexity of computing the subgame-perfect Nash equilibrium in this setting.

The rest of the paper is organized as follows. We first describe the features of our model that are common to the simultaneous and sequential setting, and provide a brief overview of the game-theoretic notions used in this paper (Sec. 2). Sec. 3 is devoted to the simultaneous voting model, while Sec. 4 focuses on the sequential voting model. We then compare the properties of the two models in Sec. 5. Sec. 6 provides an overview of the related work, and Sec. 7 concludes.

## 2 Preliminaries

### 2.1 The Basic Setup

Throughout the paper, we consider elections with $n$ *voters* $V = \{1, \ldots, n\}$ and $m$ *candidates* $C = \{c_1, \ldots, c_m\}$. We will use plurality voting as our preference aggregation rule; thus, each voter only needs to indicate a single candidate that he wants to support. Moreover, the voters are allowed to abstain. Formally, in an election each voter $i$ submits a *ballot* $b_i \in C \cup \{\bot\}$, where $b_i = c$ indicates that $i$ votes for $c$ and $b_i = \bot$ indicates that $i$ abstains. We let $B = C \cup \{\bot\}$ denote the space of all possible ballots. Given a list of ballots $\mathbf{b} = (b_1, \ldots, b_n) \in B^n$, for any $j = 1, \ldots, m$ we denote by $\sharp_j(\mathbf{b})$ the number of votes for a candidate $c_j \in C$. The *outcome* $o(\mathbf{b})$ of the election with the list of ballots $\mathbf{b} = (b_1, \ldots, b_n)$ is the set of all candidates that receive the maximum number of votes. Formally, we have $o(\mathbf{b}) = \{c_j \in C \mid \sharp_j(\mathbf{b}) \geq \sharp_i(\mathbf{b}) \text{ for all } c_i \in C\}$. We refer to the candidates in $o(\mathbf{b})$ as the *winners* of the election. We say that an election is *valid* if its outcome $o$ satisfies $o \neq \emptyset$. We denote by $\mathcal{O} = 2^C$ the set of possible outcomes of an election. We refer to outcomes $o \in \mathcal{O}$ with $|o| > 1$ as *ties*. If the outcome of an election is a tie $o$, then the eventual winner is selected from $o$ uniformly at random.

Traditionally, in the social choice literature it is assumed that voters' preferences are described by total orders over $C$. However, as in our setting the possible outcomes of the election are subsets of $C$, we need to endow the voters with preferences over $\mathcal{O}$. Now, since $|\mathcal{O}| = 2^m$, an explicit representation of such preferences would be exponential in $m$. Moreover, if voters are rational, the preferences over $\mathcal{O}$ are likely to have an additional structure: e.g., it is unlikely that some voter prefers $a$ to $b$, yet prefers $\{b, c\}$ to $\{a, c\}$ for some $a, b, c \in C$. To overcome these issues, we assume that the voters assign a utility to each candidate and rank all outcomes according to their expected utility, where the expectation is taken over the random coin tosses required to choose a single candidate in case of a tie (this implicitly assumes that the voters are risk-neutral). Formally, we encode the preferences of the voter $i$ by a vector $\mathbf{u}^i = (u_1^i, \ldots, u_m^i) \in (\mathbb{Z}^+ \cup \{0\})^m$, which assigns a *utility* $u_j^i$ to a candidate $c_j$. We interpret $u_j^i$ as the utility that $i$ obtains if eventually (i.e., after the tie between the winners is resolved) $j$ becomes the winner of the election. Consequently, we assume that given two outcomes $o_1, o_2 \in \mathcal{O} \setminus \{\emptyset\}$, the voter $i$ prefers $o_1$ to $o_2$ if and only if $u^i(o_1) > u^i(o_2)$, where for any $o \in \mathcal{O} \setminus \{\emptyset\}$ we set $u^i(o) = \frac{1}{|o|} \sum_{c_j \in o} u_j^i$. When $o$ consists of a single candidate $c_j$, to improve readability, we sometimes write $u^i(c_j)$ instead of $u^i(\{c_j\})$. We require that for each $i = 1, \ldots, n$ the vector $\mathbf{u}^i$ satisfies $u^i(o_1) \neq u^i(o_2)$ for any $o_1, o_2 \in \mathcal{O} \setminus \{\emptyset\}$ such that $o_1 \neq o_2$, i.e., no voter is indifferent between two distinct outcomes. Finally, we set $u_i(\emptyset) = -\infty$ for all $i \in V$. Note that if the values of $u^i$ are not too large, this representation is polynomial in the number of candidates. We refer to the collection of utility vectors $(\mathbf{u}^1, \ldots, \mathbf{u}^n)$ as a *preference profile*. Throughout the paper, we assume that the voters' utility vectors are common knowledge. We emphasize that we do not treat the voter's utilities as objective valuations of the usefulness of each outcome, and, in particular, we are not interested in outcomes that maximize the total utility; neither do we compare different voters' utilities. Rather, the utility vectors provide a convenient vehicle for succinctly representing the voters' preferences over ties.

Another important attribute of our model is the possibility of abstention. Informally, we assume that the voters are lazy, i.e., they vote only if this can improve the outcome from their perspective. That is, for each voter there is a small cost $\varepsilon$ associated with voting $b_i \neq \bot$, and the total payoff a voter $i$ obtains from an outcome $o$ is $u_i(o)$ if $b_i = \bot$ and $u_i(o) - \varepsilon$ if $b_i \neq \bot$. We require that $\varepsilon < \min\{|u_i(o_1) - u_i(o_2)| \mid i \in V, o_1, o_2 \in \mathcal{O} \setminus \{\emptyset\}\}$, i.e., any positive change in outcome, no matter

how small, is a sufficient motivation to vote.

We postpone describing the formal model of the voters' decision-making process till Sections 3 and 4, as it depends on whether the voters cast their ballots simultaneously (Section 3) or sequentially (Section 4).

## 2.2 Game-theoretic Solution Concepts

In this section, we provide a brief overview of some important concepts of game theory used in this paper; for a more complete treatment, the reader is referred to [10].

A *matrix game* $G$ is given by (1) a finite set of *players* $N = \{1, \ldots, n\}$; (2) for each player $i \in N$, a finite set of *actions* $A_i$; (3) a collection of *payoff functions* $p_i : A_1 \times \cdots \times A_n \to \mathbb{R}^n$, $i \in N$, where $p_i(a_1, \ldots, a_n)$ indicates the payoff of player $i$ when for each $j \in N$ player $j$ chooses the action $a_j$. It is assumed the players know each others' payoff functions, but they choose their actions simultaneously, so they cannot condition their choice of action on other players' actions. A vector of actions $(a_1, \ldots, a_n)$ is called a *(pure) Nash equilibrium (PNE)* of $G$ if $p_i(a_1, \ldots, a_i, \ldots, a_n) \geq p_i(a_1, \ldots, a_i', \ldots, a_n)$ for all $i \in N$ and all $a_i' \in A_i$. Informally, in a Nash equilibrium no agent can benefit from deviating from his prescribed action assuming that all other agents do not deviate either.

Not every matrix game has a pure Nash equilibrium. However, if we expand the space of strategies available to the players by allowing each player to choose (simultaneously and independently) a probability distribution over actions (a so-called *mixed strategy*), then there exists a strategy vector such that no player can increase his *expected* payoff by deviating. Such strategy vector is called a *mixed* Nash equilibrium; the fact that each matrix game has a mixed Nash equilibrium was proved by Nash in his seminal paper [12]. As in this paper we focus on pure Nash equilibria, we will skip the formal definition of mixed strategies and mixed Nash equilibria in this overview.

In *extensive-form* games, players take turns choosing their actions. Such games are called *perfect information* games if each player can observe all actions chosen so far. The scenario considered in this paper (Section 4) corresponds to an extensive-form perfect information game with a finite number of players in which each player moves exactly once. We will refer to such games as *sequential* games, and adapt the definitions that follow to this simplified setting.

Formally, a sequential game is given by a finite set of players $N = \{1, \ldots, n\}$, a finite set of actions $A_i$ for each player $i \in N$, and a collection of *payoff functions* $p_i : A_1 \times \cdots \times A_n \to \mathbb{R}^n$, $i \in N$. It is assumed that player 1 chooses his action $a_1$ first, and his choice is observed by all players. Then player 2 chooses his action $a_2$; again, this is observed by everyone. The process continues until the last player chooses his action $a_n$. The payoff to player $i$ is given by $p_i(a_1, \ldots, a_n)$. In such games, a *strategy* of a player $i$ is a function $s_i : A_1 \times \cdots \times A_{i-1} \to A_i$, where $s_i(a_1, \ldots, a_{i-1})$ is the action that player $i$ should take provided that for $j = 1, \ldots, i-1$ player $j$ took action $a_j$. An extensive-form game can be converted to a matrix game, by mapping the players' strategies in the sequential game to actions in the matrix game. Note that the number of actions available to player $i$ in the resulting game is doubly exponential in $i$. A vector of strategies $(s_1, \ldots, s_n)$ is a (pure) *Nash equilibrium* in the sequential game if the corresponding action profile in the matrix game is a (pure) Nash equilibrium in that game. However, in sequential games, not all pure Nash equilibria provide a good prediction of the outcome of the game, as some of them may involve non-credible threats (see [16]).

A more refined solution concept for sequential games, which will be used throughout this paper, is that of subgame-perfect Nash equilibrium (SPNE). Formally, for each $0 \leq j < n$ and each vector of actions $(a_1, \ldots, a_j)$, we can define a *subgame* $G_j(a_1, \ldots, a_j)$ with a set of players $\{j+1, \ldots, n\}$, a set of actions $A_i$ for each player $i = j+1, \ldots, n$, and payoff functions $p_i^j : A^n \to \mathbb{R}^{n-j}$, $i = j+1, \ldots, n$, given by $p_i^j(a_{j+1}, \ldots, a_n) = p_i(a_1, \ldots, a_n)$. A strategy profile $(s_1, \ldots, s_n)$ is called a *subgame-perfect Nash equilibrium (SPNE)* if for each $j = 0, \ldots, n-1$ the vector $(s_{j+1}, \ldots, s_n)$ is a Nash equilibrium in the matrix game that corresponds to $G_j(a_1, \ldots, a_j)$. It is well known that any sequential game has an SPNE in pure strategies (i.e., without randomization).

## 3 Simultaneous Voting

In simultaneous elections, all voters have to submit their ballots at the same time. As in this paper we are interested in studying the behavior of rational voters, and we assume that the voters know each others' preferences, it is natural to model this scenario as a matrix game with $n$ players and $m+1$ actions per player. As mentioned in Section 2, any such game has an equilibrium in mixed strategies. However, mixed strategies are rather unnatural in the context of voting. Thus, it would be desirable to characterize the preference profiles that correspond to pure Nash equilibria (PNE). In the rest of this section, we provide such a characterization, and study whether it leads to an efficient algorithm for checking the existence of PNE.

Our characterization relies on a sequence of lemmas.

**Lemma 1.** *For any list of ballots $\mathbf{b} \in B^n$ that constitutes a PNE and any $j, j' \in \{1, \ldots, m\}$, if $\sharp_j(\mathbf{b}) > 0$ and $\sharp_{j'}(\mathbf{b}) > 0$, then $\sharp_j(\mathbf{b}) = \sharp_{j'}(\mathbf{b})$.*

*Proof.* Suppose that there exist some $j, j' \in \{1, \ldots, m\}$ such that $\sharp_j(\mathbf{b}) > \sharp_{j'}(\mathbf{b}) > 0$. Then $c_{j'}$ is not among the winners of the election. As $\sharp_{j'}(\mathbf{b}) > 0$, there exists at least one voter $i$ such that $b_i = c_{j'}$. However, $i$ can improve his payoff by voting $\perp$ instead, as this will result in the same outcome. Hence, $\mathbf{b}$ is not a PNE, a contradiction. □

**Lemma 2.** *For any list of ballots $\mathbf{b} \in B^n$ that constitutes a PNE, we have $o(\mathbf{b}) \neq \emptyset$.*

*Proof.* Clearly, if $o(\mathbf{b}) = \emptyset$, any voter $i$ can profitably deviate from $\mathbf{b}$ by voting for his favorite candidate $c$. This will increase $i$'s utility, as $u^i(c) - \varepsilon > -\infty$. □

**Lemma 3.** *For any list of ballots $\mathbf{b} \in B^n$ that constitutes a PNE, if $o(\mathbf{b}) = \{c_j\}$ for some $c_j \in C$, then we have $\sharp_j(\mathbf{b}) = 1$, $\sharp_{j'}(\mathbf{b}) = 0$ for all $j' \neq j$.*

*Proof.* If there are several voters voting for $c_j$, one of these voters can improve his utility by abstaining: this will not change the outcome of the election, yet increase his payoff by $\varepsilon$. Thus, $\sharp_j(\mathbf{b}) = 1$. On the other hand, if $\sharp_{j'}(\mathbf{b}) > 0$ for some $j' \neq j$, then by Lemma 1 candidate $j'$ would receive as many votes as candidate $j$, so we would have $j' \in o(\mathbf{b})$. □

We are now ready to characterize the voters' preferences for which the election has a unique winner. Perhaps surprisingly, this can only happen if there is a consensus among the voters.

**Theorem 1.** *For any candidate $c_j \in C$, there exists a list of ballots $\mathbf{b} \in B^n$ with $o(\mathbf{b}) = \{c_j\}$ that constitutes a PNE if and only if all voters rank $c_j$ first, i.e., for all $i \in V$ we have $u_j^i \geq u_{j'}^i$ for all $j' \neq j$.*

*Proof.* Suppose that all voters rank $c_j$ first, and $i$ is the unique voter voting for $c_j$. Clearly, $i$ does not want to deviate: he does not benefit from voting for another candidate, and if he abstains, his payoff is $-\infty$. For any other voter, $\{c_j\}$ is their most preferred outcome, so they do not want to deviate either.

Conversely, suppose that $o(\mathbf{b}) = \{c_j\}$. By Lemma 3, we have $\sharp_j(\mathbf{b}) = 1$. Clearly, the unique voter $i$ with $b_i = c_j$ must prefer $c_j$ to all other candidates. Further, if there exists a voter $k \neq i$ such that $u_j^k < u_{j'}^k$, he can deviate by voting for $c_{j'}$ and changing the outcome from $o = \{c_j\}$ to $o' = \{c_j, c_{j'}\}$. This improves his payoff from $u^k(o) = u_j^k$ to $u^k(o') - \varepsilon = \frac{u_j^k + u_{j'}^k}{2} - \varepsilon$; by our assumption on $\varepsilon$, this change is positive. □

To describe the preferences that result in ties, we first need a technical lemma.

**Lemma 4.** *For any list of ballots $\mathbf{b} \in B^n$ that constitutes a PNE, if $|o(\mathbf{b})| > 1$, then no voter abstains, and each candidate $c \in o(\mathbf{b})$ gets exactly $\frac{n}{|o(\mathbf{b})|}$ votes.*

*Proof.* Let $C^* = o(\mathbf{b})$. Suppose that $b_i = \bot$ for some $i \in V$, and let $c_j$ be $i$'s most preferred candidate in $C^*$. Since by Lemma 1 all candidates in $C^*$ get the same number of votes, $i$ can change the outcome from $C^*$ to $c_j$ by voting for $c_j$, a contradiction with $\mathbf{b}$ being a PNE. Hence, no voter abstains, so by Lemma 1 all candidates in $C^*$ get $\frac{n}{|o(\mathbf{b})|}$ votes. □

We can now characterize the preferences that lead to ties among several candidates. Essentially, our characterization says that the outcome can be a tie if, when we restrict the voters' preferences to these candidates, each candidate gets the same number of top votes (i.e., is ranked first the same number of times).

**Theorem 2.** *For any $C^* = \{c_{j_1}, \ldots, c_{j_k}\} \subseteq C$ with $k > 1$, there exists a list of ballots $\mathbf{b} \in B^n$ with $o(\mathbf{b}) = C^*$ that constitutes a PNE if and only if $V_{j_\ell} = \frac{n}{k}$ for all $\ell \in \{1, \ldots, k\}$, where*

$$V_{j_\ell} = \{i \in V \mid u_{j_\ell}^i > u_{j_t}^i \text{ for all } c_{j_t} \in C^* \setminus \{c_{j_\ell}\}\},$$

*and, moreover, for all $\ell = 1, \ldots, k$, each voter in $V_{j_\ell}$ prefers the outcome $C^*$ to any outcome of the form $\{c_{j_t}\}$, where $c_{j_t} \in C^*$ and $t \neq \ell$.*

*Proof.* Without loss of generality, we can assume that $c_{j_\ell} = c_\ell$ for $\ell = 1, \ldots, k$, i.e., $C^* = \{c_1, \ldots, c_k\}$.

For the "if" direction, suppose that we have $|V_j| = \frac{n}{k}$ for all $j = 1, \ldots, k$, and all voters in $V_j$ prefer $C^*$ to the outcome in which some candidate $c_\ell \in C^*$, $\ell \neq j$, is the unique winner. Then the list of ballots $\mathbf{b}$ in which each voter in $V_j$ votes for $c_j$ constitutes a PNE, and $o(\mathbf{b}) = C^*$. Indeed, consider any voter $i \in V_j$. If he abstains, the outcome changes from $C^*$ to $C^* \setminus \{c_j\}$, leading to a negative change in $i$'s payoff. To complete the argument, we need to consider two cases. First, suppose that $k = n$, i.e., $C^* = C$. In this case, if $i$ changes his vote to $c_\ell$ for some $\ell \neq j$, then $c_\ell$ gets two votes, and therefore becomes the unique winner. Thus, for $k = n$, voting for $c_\ell \neq c_j$ is not beneficial for $i$. Next, suppose that $k < n$. Since $|V_j| = \frac{n}{k}$, it follows that $\frac{n}{k}$ is an integer, and therefore $|V_\ell| \geq 2$ for all $\ell = 1, \ldots, k$. Now, if $i$ changes his vote to $c_\ell$ for some $c_\ell \in C^*$, the candidate $c_\ell$ becomes the unique winner. By our assumption on voters' utilities, this lowers $i$'s payoff. Finally, if $i$ changes his vote to $c_\ell$ for some $c_\ell \notin C^*$, then $c_\ell$ gets one vote, $c_j$ gets $\frac{n}{k} - 1$ votes, and all candidates in $C^* \setminus \{c_j\}$ get $\frac{n}{k}$ votes, so the outcome is $C^* \setminus \{c_j\}$. Clearly, $u_i(C^*) > u_i(C^* \setminus \{c_j\})$, so $i$ cannot benefit from voting in this way either.

For the "only if" direction, suppose that **b** is a PNE with $o(\mathbf{b}) = C^*$. By Lemma 4, each candidate in $C^*$ gets exactly $\frac{n}{k}$ votes. Suppose that $|V_j| < \frac{n}{k}$ for some $j = 1, \ldots, k$. Then not all voters who vote for $c_j$ in **b** prefer $c_j$ to all other candidates in $C^*$. Thus, any of these voters can achieve a higher payoff by voting instead for his favorite candidate in $C^*$, as that would make that candidate the unique winner. Thus, we have that $V_j = \frac{n}{k}$ for all $j = 1, \ldots, k$. Now, suppose there is some $j \in \{1, \ldots, k\}$ and some $i \in V_j$ such that $i$ prefers an outcome of the form $\{c_\ell\}$ to $C^*$ for some $c_\ell \in C^* \setminus \{c_j\}$. Then $i$ can make $c_\ell$ the unique winner by changing her vote to $c_\ell$, a contradiction. □

Observe that Theorem 2 implies that the outcome of an election can be a tie with $k$ winners only if $n$ is divisible by $k$. Thus, for example, if $n$ is a prime number, there are no PNE for a given preference profile unless all voters rank the same candidate first.

We can use Theorems 1 and 2 to show that even when a PNE exists, it is not necessarily unique.

**Example 1.** Consider an election with three candidates $A, B, C$ and four voters whose utilities are given by $u^1(A) = u^2(A) = u^3(A) = u^4(A) = 5$, $u^1(B) = u^2(B) = 2$, $u^1(C) = u^2(C) = 1$, $u^3(B) = u^4(B) = 1$, $u^3(C) = u^4(C) = 2$. This election has an obvious PNE with $\mathbf{b} = (A, \bot, \bot, \bot)$. However, as Theorem 2 predicts, it also has another PNE, namely, $\mathbf{b}' = (B, B, C, C)$. Intuitively, $\mathbf{b}'$ is a PNE because no voter can unilaterally change the outcome to $A$. By adding $m - 3$ dummy candidates that are ranked below $A$ but above $B$ and $C$ by all voters, we conclude that the winners in PNE can be candidates that are not ranked in top $m - 2$ positions by *any* voter.

Clearly, not all preference profiles satisfy the criterion put forward in the statement of Theorem 2. Thus, a natural question to ask is whether we can check that a given preference profile satisfies this criterion. For a fixed set $C^*$, such a check is easy to perform. However, it turns out that checking whether a profile $(\mathbf{u}^1, \ldots, \mathbf{u}^n)$ satisfies this criterion for *some* set $C^*$ is NP-complete.

**Theorem 3.** *Given a set of voters $V$ with $|V| = n$, a set of candidates $C$ with $|C| = m$ and a collection of utility vectors $(\mathbf{u}^1, \ldots, \mathbf{u}^n)$, it is* NP-*complete to check whether there exists a ballot vector $\mathbf{b} = (b_1, \ldots, b_n)$ that constitutes a PNE for this election.*

*Proof.* It is easy to see that the problem is in NP. Indeed, we can first check if there is a candidate that is ranked first by all voters. If this is the case, by Theorem 1, there is PNE in which this candidate wins. Otherwise, we can guess the set of winners $C^*$ and check that $C^*$ satisfies the criterion of Theorem 2.

To show NP-hardness, we reduce from the classic NP-hard problem EXACT COVER BY 3-SETS (X3C). An instance of X3C is given by a ground set $G = \{g_1, \ldots, g_N\}$, and a collection of feasible sets $\mathcal{E} = \{E_1, \ldots, E_M\}$, where $E_i \subset G$ and $|E_i| = 3$ for all $i = 1, \ldots, M$. It is a "yes"-instance if $G$ can be covered by exactly $\frac{N}{3}$ sets from $\mathcal{E}$, and a "no"-instance otherwise.

Given an instance $(G, \mathcal{E})$ of X3C, for each $g \in G$ let $f(g)$ be the *thickness of the cover* at $g$, i.e., set $f(g) = |\{E_j \in \mathcal{E} \mid g \in E_j\}|$. Let us say that an instance of X3C is *uniform* if each element of the ground set is contained in the same number of feasible sets, i.e., there exists some $f > 0$ such that $f(g) = f$ for all $g \in G$.

Now, suppose that we are given an instance $(G, \mathcal{E})$ of X3C. Note that we can assume without loss of generality that this instance is non-uniform. Indeed, if $f(g) = 1$ for all $g \in G$, then $(G, \mathcal{E})$ is obviously a "yes"-instance with the set $\mathcal{E}$ itself being an exact cover, so in our reduction we can output an arbitrary "yes"-instance of our problem and stop. Otherwise, if $f(g) = f > 1$ for all $g \in G$, we can modify $(G, \mathcal{E})$ by adding three new elements $x, y, z$ to $G$ and a new feasible set $\{x, y, z\}$ to $\mathcal{E}$. Clearly, the original instance is a "yes"-instance if and only if the new one is, and the new instance is non-uniform. Renumber the elements of the ground set in order of non-increasing values of $f(g)$, i.e., so that $f(g_1) \geq f(g)$ for all $g \in G$.

We will now construct an instance of our problem as follows. Our instance will contain $m = N + M$ candidates and $n = 2N + 3M$ voters. Specifically, we create a candidate $d_i$ for each $g_i \in G$ and a candidate $e_j$ for each $E_j \in \mathcal{E}$. Set $C_d = \{d_i \mid i = 1, \ldots, N\}$, $C_e = \{e_j \mid j = 1, \ldots, M\}$, and let $C = C_d \cup C_e$.

We will now describe the voters' preferences over these candidates. In our construction, each voter is characterized by his preference ranking over the candidates, and the utility that a voter $i$ assigns to a candidate $j$ can be determined as a function of $j$'s position in $i$'s rankings. Specifically, each voter assigns a utility of $(m+1)^{(m-j)}$ to his $j$-th most preferred candidate. It is not hard to check that given such utility vectors, no voter assigns the same utility to two different outcomes, and for any set of alternatives $C^*$, any voter $i$ prefers $C^*$ to the outcome in which $i$'s second most favorite alternative in $C^*$ is the unique winner. It remains to describe how the voters ranks the candidates. For each $g_i \in G$, there are two voters $u_{(i,1)}$ and $u_{(i,2)}$ that rank $d_i$ first, followed by all other candidates in $C_d$ in lexicographic order, followed by all candidates in $C_e$ in lexicographic order. Let $U$ denote the set of all such voters. Further, for each $E_j = \{g_x, g_y, g_z\} \in \mathcal{E}$, we construct three voters $w_{(j,x)}$, $w_{(j,y)}$, and $w_{(j,z)}$. For each $t \in \{x, y, z\}$, the voter $w_{(j,t)}$ ranks $e_j$ first, followed by $d_t$, followed by all candidates in $C_d \setminus \{d_t\}$ in lexicographic order, followed by all candidates in $C_e \setminus \{e_j\}$ in lexicographic order. Let $W$ denote the set of all such voters, and set $V = U \cup W$.

Note that in the resulting instance each candidate in $C_d$ gets two top votes, and each candidate in $C_e$ gets three top votes. Note also that according to Theorem 2, and given the shape of the voters' utilities, our problem can now be restated as follows: is it possible to delete at most $m - 1$ candidates from $C$ so that all surviving candidates get the same number of top votes in the preference profiles restricted to the set of survivors?

We are now ready to prove the correctness of our reduction. Suppose that the input instance of X3C is a "yes"-instance with an exact cover $\mathcal{F} \subset \mathcal{E}$. Then the resulting instance of our problem is a "yes"-instance, too: it suffices to delete the candidates in $C_e$ that correspond to sets in $\mathcal{F}$. Indeed, after there candidates are deleted, each candidate in $C_d$ gets exactly one extra vote (specifically, if $g_i$ is covered by $E_j$ in $\mathcal{F}$, candidate $d_i$ gets an extra vote from $w_{(j,i)}$), and hence all surviving candidates get exactly three votes.

We will now argue that if deleted candidates do not correspond to an exact cover of $G$, then in the resulting elections not all candidates get the same number of top votes. Denote the set of surviving candidates by $S$.

We will consider three cases.

- $C_d \cap S = \emptyset$, $C_e \cap S \neq \emptyset$. Let $e_j$ be the first surviving candidate from $C_e$, i.e., $j = \min_{e_k \in S} k$. Then $e_j$ gets at least $2N + 3$ votes (in particular, $e_j$ is now the top choice of all voters in $U$), while all other candidates in $S$ get exactly 3 votes.

- $C_d \cap S \neq \emptyset$, $C_e \cap S = \emptyset$. If $C_d = S$, the statement follows from our assumption that the input instance of X3C is non-uniform. In particular, $d_1$ gets at least as many votes as any other candidate (and more votes than $d_N$). Otherwise, let us start with $S' = C_d$ and delete candidates from $S'$ one by one so as to obtain $S$. If $d_1 \in S$, each such step increases the number of top votes that $d_1$ receives by at least two (namely, he now receives the votes of the voters in $U$ that ranked the deleted candidate first, and maybe also some votes in $W$), and does not increase the number of top votes that any other candidate receives. Thus, in $S$ candidate $d_1$ gets more votes than any other candidate. If $d_1 \notin S$, let $d_j$ be the first surviving candidate from $C_d$, i.e., $j = \min_{d_k \in S} k$. Modify our deletion procedure so that it starts with $S' = C_d$, first deletes all candidates $d_k \in C_d \setminus S$ with $k > j$, and then deletes $d_1, \ldots, d_{j-1}$. After the first stage of this procedure, $d_1$ has strictly more votes than any candidate that survives at this point. When $d_1$ gets deleted, all his top votes move to $d_2$, so $d_2$ becomes the unique winner. In the end, $d_j$ gets all votes that $d_1$ had at the end of the first stage, so he is the unique winner in $S$.

- $C_d \cap S \neq \emptyset$, $C_e \cap S \neq \emptyset$. Note that in this case each candidate in $C_e \cap S$ gets exactly three votes. Now, if $C_d \not\subset S$, the first surviving candidate in $C_d$ gets at least four votes (he is now the top choice of all voters in $U$ that ranked the candidates from $C_d \setminus S$ first), so the final outcome is not a tie. Thus, we can assume that $C_d \subset S$. Therefore, for the outcome to be a tie among all candidates in $S$, it has to be the case that each candidate $d_i \in C_d$ gets exactly one vote in addition to $u_{(i,1)}$ and $u_{(i,2)}$. But this means that the sets that correspond to $C_e \cup (C \setminus S)$ form an exact cover of $G$.

$\square$

We remark that if the number of candidates or the number of voters is small, checking if a pure Nash equilibrium exists (and constructing one, if this is the case) is easy.

**Proposition 1.** *Suppose that we are given a simultaneous election with the set of voters $V$, $|V| = n$, the set of candidates $C$, $|C| = m$, and a collection of utility vectors $\mathbf{u}^i$, $i = 1, \ldots, n$. Set $u_{\max} = \max\{u_j^i \mid i \in V, j = 1, \ldots, m\}$. Then we can check if this election has a PNE and construct one if it exists in time $O(nm2^m \log u_{\max})$. We can also solve this problem in time $O(m^2 n(m+1)^n \log u_{\max})$.*

*Proof.* It is easy to design an algorithm whose running time is bounded by $O(nm2^m \log u_{\max})$: we can simply consider all subsets of candidates and check if any of them satisfies the criterion stated in Theorem 2. For each such subset, this check can be performed in time $O(nm \log u_{\max})$. Of course, we also check if there is a candidate that is ranked first by every voter, to see whether the election satisfies the criterion stated in Theorem 1.

Alternatively, we can simply consider all $(m+1)^n$ possible action profiles and check whether any of them is a PNE. For a fixed profile, we need to consider $m$ possible deviations for each of the $n$ players, and compute the corresponding utilities, so the running time of this algorithm is bounded by $O(m^2 n(m+1)^n \log u_{\max})$. $\square$

## 4 Sequential Voting

In sequential elections, the voters submit their votes one by one, in a predetermined order. Formally, a sequential election is described by a permutation $\pi : \{1, \ldots, n\} \to \{1, \ldots, n\}$ and consists of $n$ rounds. During the $i$th round of an election $\pi$, voter $\pi(i)$ submits a ballot $b_{\pi(i)}$ in $B$. We set $r_i = b_{\pi(i)}$ for all $i \in V$ and refer to the vector $\mathbf{r} = (r_1, \ldots, r_n)$ as the *vote vector*. We overload the notation by using $\sharp_j(\mathbf{r})$ and $o(\mathbf{r})$ to denote the number of appearances of $c_j$ in $\mathbf{r}$ and the outcome that corresponds to $\mathbf{r}$, respectively.

We will now describe how the voters decide which candidates to support. Each voter knows the ordering $\pi$ as well as the preferences of all voters. We also assume that each voter $\pi(i)$ can observe the votes of all players in rounds $1, \ldots, i-1$. As we are interested in voters that are rational expected utility maximizers, this setting can be viewed as a sequential game, and an appropriate solution concept for it is subgame-perfect Nash equilibrium (SPNE). As argued in Section 2, each finite sequential game with perfect information possesses an SPNE. Further, since we assume that the players' utilities for all outcomes are distinct, all SPNE of a given election correspond to the same outcome. Therefore, it is not important which of the SPNE of a given election is selected, and hence we can assume that the players act according to some fixed SPNE (e.g., one that is minimal with respect to some fixed ordering over the SPNE of the given game).

The players' strategies in this SPNE can be described as functions of the history of the game. Specifically, given a player $j$ with $j = \pi(i)$, his strategy can be described as a function $f_j : B^{i-1} \to B$. Note also that $f_{\pi(i)}$ does not depend on the identities of the voters that have cast a particular vote in rounds $1, \ldots, i-1$, but only on the number of previous votes for each of the candidates.

To build some intuition for the voters' behavior in sequential voting, we start with the case where there are two candidates only. We then explore which of our conclusions for the 2-candidate setting apply in the general case.

## 4.1 Two Candidates

In this section, we focus on the case of two candidates; we denote these candidates by $A$ and $B$. In this case, each voter's preference relation over outcomes is uniquely determined by whether he prefers $A$ to $B$ or vice versa: indeed, we have $u^i(\{A, B\}) = \frac{u^i(\{A\}) + u^i(\{B\})}{2}$, so $u^i(\{A\}) > u^i(\{B\})$ implies $u^i(\{A\}) > u^i(\{A, B\}) > u^i(\{B\})$ and $u^i(\{A\}) < u^i(\{B\})$ implies $u^i(\{A\}) < u^i(\{A, B\}) < u^i(\{B\})$. We will therefore refer to voters that prefer $A$ to $B$ as $A$-voters, while the voters that prefer $B$ to $A$ will be referred to as $B$-voters. Now, any election can be concisely represented as a string in $\{A, B\}^n$, where the $i$th component of this string indicates whether the voter $\pi(i)$ is an $A$-voter or a $B$-voter.

We start by considering two simple examples.

**Example 2.** Consider an election that corresponds to the string $ABABA$, i.e., the even-numbered voters prefer $B$, and the odd-numbered voters prefer $A$. We claim that the unique vote vector that corresponds to the SPNE of this election is $(A, \bot, \bot, \bot, \bot)$. Indeed, if the first voter does not vote, the $B$-voters can ensure that the outcome is either $\{A, B\}$ or $\{B\}$, so the first voter prefers to vote for $A$. After that, if the first of the $B$-voters chose to vote, the remaining two $A$-voters could nevertheless ensure that $A$ remains the unique winner, so the first $B$-voter prefers to abstain. Therefore, the second $A$-voter does not need to vote: indeed, even if he stays home and the remaining $B$-voter goes to the polls, the last $A$-voter can still make $A$ the unique winner. Now, the second $B$-voter prefers not to vote since his vote would be canceled by the vote of the last $A$-voter. Therefore, the last $A$-voter does not need to vote either.

Clearly, this example generalizes to the case where there are $k+1$ $A$-voters, $k$ $B$-voters, and $A$-voters and $B$-voters alternate, starting with an $A$-voter. The analysis above shows that in any such setting the vote vector will be of the form $(A, \bot, \ldots, \bot)$.

Our second example shows that, while reordering voters does not change the outcome of the election (for a proof of this statement for any 2-candidate election, see Corollary 1), it can drastically change the number of votes obtained by the winner.

**Example 3.** Again, consider an election with $k+1$ $A$-voters, and $k$ $B$-voters, where the $A$-voters vote first, followed by the $B$-voters. In this case, if any of the $A$-voters chooses to abstain or to vote for $B$, the $B$-voters can ensure a tie between $A$ and $B$, or, if more than one $A$-voter abstains, make $B$ the unique winner. Thus, all $A$-voters vote for $A$. After this, the $B$-voters cannot change the outcome of the election, so they all abstain, and the vote vector is $(A, \ldots, A, \bot, \ldots, \bot)$, i.e., $A$ obtains $k+1$ votes.

Now, suppose that all $B$-voters vote first, followed by the $A$-voters. In this case, the $B$-voters know that no matter how they vote, the $A$-voters can still make $A$ the unique winner, so they stay home. Further, each of the first $k$ $A$-voters knows that he does not need to vote, as he can count on the last voter to vote for $A$. Thus, in this case the vote vector is $(\bot, \ldots, \bot, A)$, i.e., $A$ obtains one vote only, just like in the case of alternating voters.

In general, it turns out that for two candidates our game has a unique SPNE, and the voters' SPNE strategies have a simple intuitive description. Namely, each voter "predicts" the total number of votes for each $X \in \{A, B\}$ by adding up the number of votes already cast for $X$ and the number of future $X$-voters. He then votes for his preferred candidate if the predicted outcome is a tie (i.e., by voting he can make his candidate the unique winner), or if his candidate trails the other candidate by one vote (i.e., by voting he can turn the outcome into a tie). We will now prove that this simple heuristic results in an SPNE.

**Theorem 4.** *Consider an election with $n$ voters and two candidates $A$ and $B$ that is described by a permutation $\pi$. For $i = 1, \ldots, n$ and $X \in \{A, B\}$, let $p(i, X)$ denote the number of votes received by $X$ prior to round $i$, and let $f(i, X)$ denote the number of $X$-voters in rounds $i+1, \ldots, n$. Set also $\overline{A} = B$, $\overline{B} = A$. Let $\mathbf{r}$ be the vote vector that corresponds to an SPNE of this game. Then for any $i = 1, \ldots, n$ and $X \in \{A, B\}$, if $\pi(i)$ is an $X$-voter, then $r_{\pi(i)} = X$ if and only if*

$$p(i, X) + f(i, X) \in \{p(i, \overline{X}) + f(i, \overline{X}), p(i, \overline{X}) + f(i, \overline{X}) - 1\},$$

*and $r_{\pi(i)} = \bot$ in all other cases. Further, for any $i = 1, \ldots, n$ and $X \in \{A, B\}$, if voter $i$ is an $X$-voter, then we have*

$$o(\mathbf{r}) = \begin{cases} \{X\} & \text{if } p(i, X) + f(i, X) > p(i, \overline{X}) + f(i, \overline{X}) - 1, \\ \{A, B\} & \text{if } p(i, X) + f(i, X) = p(i, \overline{X}) + f(i, \overline{X}) - 1, \\ \{\overline{X}\} & \text{if } p(i, X) + f(i, X) < p(i, \overline{X}) + f(i, \overline{X}) - 1. \end{cases}$$

*Proof.* We can assume without loss of generality that $\pi(i) = i$ for $i = 1, \ldots, n$.

The proof is by backwards induction. The statement is clearly true for $i = n$. Indeed, if $n$ is an $A$-voter, he chooses to vote if and only if before he votes, either $A$ and $B$ have the same number of votes (and hence $n$ can make $A$ the unique winner) or $B$ leads by one vote (and hence $n$ can turn the outcome into a tie between $A$ and $B$). In all other cases, $n$ cannot improve the election outcome.

Now, suppose that the statement of the theorem is true for $i+1$, and consider voter $i$. We can assume without loss of generality that $i$ is an $A$-voter. We need to consider two cases that depend on the preferences of the next voter.

- Voter $i+1$ is an $A$-voter. In this case, if $i$ votes for $A$, we obtain

$$\begin{aligned} p(i+1, A) &= p(i, A) + 1, & f(i+1, A) &= f(i, A) - 1, \\ p(i+1, B) &= p(i, B), & f(i+1, B) &= f(i, B). \end{aligned}$$

Similarly, if $i$ votes for $B$, we obtain

$$p(i+1, A) = p(i, A), \qquad f(i+1, A) = f(i, A) - 1,$$
$$p(i+1, B) = p(i, B) + 1, \qquad f(i+1, B) = f(i, B).$$

Finally, if $i$ abstains, we obtain

$$p(i+1, A) = p(i, A), \qquad f(i+1, A) = f(i, A) - 1,$$
$$p(i+1, B) = p(i, B), \qquad f(i+1, B) = f(i, B).$$

Suppose first that
$$p(i, A) + f(i, A) < p(i, B) + f(i, B) - 1.$$
Then it is easy to check that no matter how $i$ votes,
$$p(i+1, A) + f(i+1, A) < p(i+1, B) + f(i+1, B) - 1.$$
Thus, by inductive assumption, $B$ wins, so $i$ prefers not to vote.

Now, suppose that
$$p(i, A) + f(i, A) = p(i, B) + f(i, B) - 1.$$
If $i$ votes for $A$, we obtain
$$p(i+1, A) + f(i+1, A) = p(i+1, B) + f(i+1, B) - 1,$$
and, by inductive assumption, the outcome is $\{A, B\}$. If $i$ votes for $B$ or abstains, we obtain
$$p(i+1, A) + f(i+1, A) = p(i+1, B) + f(i+1, B) - 1,$$
i.e., $B$ wins. Thus, in this case $i$ prefers to vote for $A$, and hence the final outcome is $\{A, B\}$.

Next, suppose that
$$p(i, A) + f(i, A) = p(i, B) + f(i, B).$$
If $i$ votes for $A$, we obtain
$$p(i+1, A) + f(i+1, A) = p(i+1, B) + f(i+1, B),$$
and, by inductive assumption, the outcome is $\{A\}$. If $i$ abstains, we obtain
$$p(i+1, A) + f(i+1, A) = p(i+1, B) + f(i+1, B) - 1,$$
and hence the outcome is $\{A, B\}$. Finally, if $i$ votes for $B$, we obtain
$$p(i+1, A) + f(i+1, A) = p(i+1, B) + f(i+1, B) - 2,$$
and hence the outcome is $\{B\}$. Thus, in this case $i$ prefers to vote for $A$, and hence the final outcome is $\{A\}$.

Finally, suppose that
$$p(i, A) + f(i, A) > p(i, B) + f(i, B).$$
Then if $i$ votes for $A$ or abstains,
$$p(i+1, A) + f(i+1, A) \geq p(i+1, B) + f(i+1, B) - 1,$$
and $A$ wins. Thus, $i$ prefers to abstain, and the final outcome is $\{A\}$.

- Voter $i+1$ is a $B$-voter. In this case, if $i$ votes for $A$, we obtain

$$p(i+1, A) = p(i, A) + 1, \qquad f(i+1, A) = f(i, A),$$
$$p(i+1, B) = p(i, B), \qquad f(i+1, B) = f(i, B) - 1.$$

Similarly, if $i$ votes for $B$, we obtain

$$p(i+1, A) = p(i, A), \qquad f(i+1, A) = f(i, A),$$
$$p(i+1, B) = p(i, B) + 1, \qquad f(i+1, B) = f(i, B) - 1.$$

Finally, if $i$ abstains, we obtain

$$p(i+1, A) = p(i, A), \qquad f(i+1, A) = f(i, A),$$
$$p(i+1, B) = p(i, B), \qquad f(i+1, B) = f(i, B) - 1.$$

Suppose first that
$$p(i, A) + f(i, A) < p(i, B) + f(i, B) - 1.$$
Then no matter how $i$ votes,
$$p(i+1, A) + f(i+1, A) \le p(i+1, B) + f(i+1, B),$$
so by inductive assumption $B$ wins. Thus, $i$ prefers to abstain.

Now, suppose that
$$p(i, A) + f(i, A) = p(i, B) + f(i, B) - 1.$$
If $i$ votes for $A$, we have
$$p(i+1, A) + f(i+1, A) = p(i+1, B) + f(i+1, B) + 1,$$
so by inductive assumption the outcome is $\{A, B\}$. On the other hand, if $i$ votes for $B$ or abstains, we have
$$p(i+1, A) + f(i+1, A) \le p(i+1, B) + f(i+1, B),$$
so the outcome is $\{B\}$. Thus, $i$ prefers to vote for $A$, and the outcome is $\{A, B\}$.

Next, suppose that
$$p(i, A) + f(i, A) = p(i, B) + f(i, B).$$
If $i$ votes for $A$, we have
$$p(i+1, A) + f(i+1, A) = p(i+1, B) + f(i+1, B) + 2,$$
i.e.,
$$p(i+1, B) + f(i+1, B) < p(i+1, A) + f(i+1, A) - 1,$$
so $A$ wins. If $i$ abstains, we have
$$p(i+1, A) + f(i+1, A) = p(i+1, B) + f(i+1, B) + 1,$$

so the outcome is $\{A, B\}$. If $i$ votes for $B$, we have

$$p(i+1, A) + f(i+1, A) = p(i+1, B) + f(i+1, B) + 1,$$

so $B$ wins. Thus, $i$ votes for $A$, and $A$ wins.

Finally, suppose that

$$p(i, A) + f(i, A) > p(i, B) + f(i, B).$$

Then if $i$ votes for $A$ or abstains, we have

$$p(i+1, A) + f(i+1, A) > p(i+1, B) + f(i+1, B) + 1,$$

so $A$ wins. Thus, $A$ prefers to abstain.

□

By substituting $i = 1$ into the criterion put forward in Theorem 4, we conclude that in the two-candidate case the majority-supported candidate always wins.

**Corollary 1.** *Consider an election with two candidates $A$ and $B$. The candidate $A$ is a unique winner of the election if and only if the number of $A$-voters exceeds the number of $B$-voters. Consequently, the outcome is a tie between $A$ and $B$ if and only if the number of $A$-voters equals the number of $B$-voters.*

Further, we say that a voter is *sincere* if he either votes for his most preferred candidate or abstains. Another corollary of Theorem 4 is that in SPNE of elections with two candidates all voters are sincere.

**Corollary 2.** *In any election with two candidates, all voters are sincere. Moreover, each voter abstains if and only if his vote cannot affect the outcome of the election.*

In Section 4.2, we show that for elections with three or more candidates, the conclusions of Corollaries 1 and 2 are no longer true.

Now, Corollary 1 implies that the outcome of the election does not depend on the voters' order. However, Examples 2 and 3 demonstrate that changing the order of voters can have a significant effect on the number of votes received by the winner. This may be important, e.g., in settings where the number of voters that expressed their support for the candidate is interpreted as the "mandate" that this candidate has. For example, a candidate with large support may be able to implement plans that are good for the society in the long run, yet are not popular with the voters due to their short-term consequences. Thus, from the perspective of the election authorities, a natural question is whether one can maximize of minimize the number of votes obtained by the winner by changing the order of voters. It turns out that for the case of two candidates, we can characterize the range of "mandates" that can be obtained by rearranging the voters.

**Theorem 5.** *Consider an election with a set of voters $V = \{1, \ldots, n\}$ and two candidates $A$ and $B$. Let $n(A)$ and $n(B)$ denote the number of $A$-voters and $B$-voters, respectively. If $n(A) = n(B)$, then for any permutation $\pi : V \to V$, we have $\sharp_A(\pi) = \sharp_B(\pi)$. On the other hand, if $n(A) > n(B)$, then for any permutation $\pi : V \to V$ we have $\sharp_B(\pi) = 0$ and $1 \leq \sharp_A(\pi) \leq n(B) + 1$. Furthermore, for any $k = 1, \ldots, n(B) + 1$, there exists a permutation $\pi_k : V \to V$ such that $\sharp_A(\pi_k) = k$.*

*Proof.* Suppose first that $n(A) = n(B)$, and consider any $\pi : V \to V$. We can show by induction on $i$ that if $\pi(i)$ is an $A$-voter, we have $p(i, A) + f(i, A) = p(i, B) + f(i, B) - 1$, and if $\pi(i)$ is a $B$-voter, we have $p(i, B) + f(i, B) = p(i, A) + f(i, A) - 1$. Therefore, by Theorem 4, $\pi(i)$ will vote for his preferred candidate.

Now, suppose that $n(A) > n(B)$, and consider an arbitrary permutation $\pi : V \to V$. By applying Theorem 4 with $i = 1$, we conclude that $A$ wins the election. Thus, $\sharp_A(\pi) \geq 1$. Suppose for the sake of contradiction that $\sharp_A(\pi) > n(B) + 1$, and consider the last voter to vote for $A$. At this point, $A$ already has at least $n(B) + 1$ votes, and $B$ cannot get more than $n(B)$ points. Thus, the current voter's vote is not needed for $A$ to win, so he should have abstained, a contradiction. Finally, suppose that some $B$-voter $\pi(j)$ does not abstain (and hence by Corollary 2 votes for $B$). Then $\pi(j)$ can profitably deviate by abstaining. Indeed, the current outcome $o = \{A\}$ is his least preferred valid outcome. Thus he can improve his payoff by abstaining as long as the outcome remains valid. To see that this is always the case, observe that if $j = n$, then at least one voter has already voted for $A$. On the other hand, if $j < n$ and no votes are cast in the first $n - 1$ rounds after $\pi(j)$ deviates, we can count on the last voter to vote for his preferred candidate.

To prove the last statement of the theorem, suppose that $n(A) > n(B)$ and fix a $k$ such that $1 \leq k \leq n(B) + 1$. We construct a permutation $\pi_k$ as follows: put $n(A) - n(B) + k - 1$ $A$-voters first, followed by all $B$-voters, followed by the remaining $n(B) - k + 1$ $A$-voters. We claim that the only voters who will not abstain are the last $k$ among the first batch of $A$-voters. This can be proved formally by induction, using Theorem 4. Informally, it is clear that the first $n(A) - n(B) - 1$ $A$-voters do not need to vote, as there is enough future $A$-voters to ensure that $A$ wins. For the next $k$ voters, it is the case that if any of them abstains, the $B$-voters can ensure that the outcome is at least a tie, so they cannot afford to abstain. Now, the $B$-voters know that if any subset of them decides to vote, the effect of their vote can be canceled by subsequent $A$-voters, so they prefer to abstain. Finally, the last $n(B) - k + 1$ $A$-voters clearly do not need to vote. □

We conclude that for two candidates sequential voting generally produces a reasonable outcome: the candidate preferred by the majority always wins and the voters behave in a very natural manner. In contrast, we will now show that even with three candidates sequential voting may produce some rather counterintuitive results.

### 4.2 Three or More Candidates

For elections with more than two candidates, we are not aware of an analogue of Theorem 4, i.e., we do not have a simple description of voters' strategies. We will now provide several examples that illustrate the voters' behavior under sequential voting.

All examples in this section are for elections with three candidates. In what follows, we denote these candidates by $A$, $B$, and $C$. We will often (partially) describe a voter by specifying his preference ordering over the candidates, and only specify the utilities that this voter assigns to specific alternatives where this is necessary for the proof. That is, we refer to a voter that prefers $A$ to $B$ to $C$ as an $ABC$-voter. Note that if we know that a voter is an $ABC$-voter, to determine his preferences over all possible outcomes, it suffices to know whether he prefers $\{A, C\}$ over $\{B\}$. Indeed, for any $o \in \mathcal{O}' = \{\{B\}, \{A, C\}, \{A, B, C\}\}$ we always have $u(\{A\}) > u(\{A, B\}) > u(o) > u(\{B, C\}) > u(\{C\})$, and knowing whether $u(\{A, C\}) > u(\{B\})$ allows us to determine the voter's preferences over the outcomes in $\mathcal{O}'$. Specifically, $u(\{A, C\}) > u(\{B\})$ implies $u(\{A, C\}) > u(\{A, B, C\}) > u(\{B\})$, and $u(\{A, C\}) < u(\{B\})$ implies $u(\{A,C\}) < u(\{A,B,C\}) < u(\{B\})$. In what follows, we refer to voters that prefer a tie between their first and third alternative to their second alternative as *top-bottom* voters, and to voters with opposite preferences as *centrist* voters.

It turns out that, unlike in the case of two candidates, the order of voters and the utilities the voters assign to alternatives play an important role in determining the outcome of the election.

**Example 4.** Consider an election with an $ABC$-voter, a $BCA$-voter, and a $CAB$-voter, voting in this order. Suppose first that all voters are top-bottom voters, so, in particular, $u_1(\{A,B,C\}) > u_1(\{B\})$, $u_2(\{B,C,A\}) > u_2(\{C\})$, $u_3(\{C,A,B\}) > u_3(\{A\})$. It is not hard to check that in this case each voter votes for his top candidate, and the outcome is a tie among all three candidates.

Now, suppose that all voters are centrist voters. In this case, if the first two voters vote for their top candidates, the last voter prefers to vote for $A$, making $A$ the unique winner. This is not acceptable to the second voter. Therefore, if the first voter votes for $A$, the second voter prefers to vote for $C$, in which case the third voter, too, votes for $C$, and $C$ becomes the unique winner. This, in turn, is not acceptable to the first voter. On the other hand, if the first voter votes for $B$, the second voter can also vote for $B$, thus making $B$ the unique winner. It is not hard to verify that $\mathbf{r} = (B, B, \bot)$ is indeed an SPNE.

This illustrates that to determine the outcome, we need to know the voters' utilities, and not just their preference orderings. Further, in our second scenario, if we reorder the voters, we can obtain a different election outcome. For example, the cyclic shift by 1, i.e., $\pi(i) = i + 1 \mod 3$ results in $A$ winning, and the cyclic shift by 2, i.e., $\pi(i) = i + 2 \mod 3$ results in $C$ winning.

Further, for three candidates it is no longer true that it is in each voter's best interest to vote sincerely. Indeed, we will now describe an election where a voter's equilibrium strategy is to vote for his *least* preferred candidate. Another interesting phenomenon that occurs in this election is that a voter abstains for strategic reasons: i.e., he prefers the outcome that arises when he abstains to any of the outcomes that can be achieved by voting for a particular candidate.

**Example 5.** Consider an election with three $ACB$-voters, two $BCA$-voter, and two $CBA$-voters, voting in this order. Suppose also that all voters are top-bottom voters. Suppose first that all three $ACB$-voters vote for $A$. Now, if the $BCA$-voters both vote for $B$, the possible outcomes are $\{A,B\}$, $\{A\}$, or $\{B\}$. Of those outcomes, the $CBA$-voters prefer $\{B\}$, so they will both vote for $B$, and the outcome is $B$. Clearly, this is the most preferred outcome for the $BCA$-voters, so voting $B$ is their SPNE strategy for the case when all $ACB$-voters vote for $A$.

Now, suppose that the first $ACB$-voter abstains, while the remaining two $ACB$-voters vote for $A$. Now, if the first $BCA$-voter votes for $A$, while the second $BCA$-voter votes for $B$ (or vice versa), the $CBA$-voters will have to vote for $B$, so the outcome is $\{A,B\}$. By checking all possible ballots for the $BCA$-voters, one can verify that any other pair of votes results in a worse outcome for these voters. Specifically, if $\{b_4, b_5\}$ is equal to $\{B,C\}$, $\{C,C\}$ or $\{C,\bot\}$, the outcome is $\{C\}$, if $\{b_4, b_5\}$ is equal to $\{A,C\}$, $\{B,\bot\}$ or $\{\bot,\bot\}$, the outcome is $\{A,C\}$, if $\{b_4, b_5\}$ is equal to $\{A,A\}$ or $\{A,\bot\}$ the outcome is $\{A\}$, and if $\{b_4, b_5\}$ is equal to $\{B,B\}$, the outcome is $\{A,B,C\}$. Thus, submitting one vote for $A$ and one vote for $B$ when two $ACB$-voters vote for $A$ is an SPNE strategy for the $BCA$-voters. Note that this means that one of the $BCA$-voters votes for his least preferred candidate, namely, $A$.

It follows that if one of the $ACB$-voters abstains, the outcome changes from $\{B\}$ to $\{A,B\}$, an improvement from that voter's perspective. Moreover, we will now argue that voting in this manner (i.e., the first $ACB$-voter abstains, and the remaining two $ACB$-voters vote for $A$) is the SPNE strategy for the first three voters. We first show that none of the $ACB$-voters can vote for $B$ or $C$. Indeed, if two of the $ACB$-voters vote for some $X \in \{B, C\}$, $X$ becomes the unique winner. Otherwise, if one of the $ACB$-voters votes for $B$, but none of them votes for $C$ (or vice versa), the outcome is $B$ (respectively, $C$). Finally, if these voters submit one vote for $B$ and one vote for $C$, the outcome is a tie between $B$ and $C$. Thus, all $ACB$-voters vote for $A$ or abstain. Further, we have already analyzed what happens if the $ACB$-voters submit three or two votes for $A$. It remains to analyze two cases: the $ACB$-voters collectively submit one vote for $A$, and all $ACB$-voters abstain. It is easy to see that in both of these cases the outcome is $\{B, C\}$.

We have shown that for two candidates, sequential voting always selects the *majority winner*, i.e., the candidate that is preferred by the majority of voters, when one exists. A natural generalization of the notion of majority winner to the case of more than two alternatives is that of *Condorcet winner*: a candidate $c$ is said to be a Condorcet winner if for any $c' \in C \setminus \{c\}$ the majority of voters prefers $c$ to $c'$. While the Condorcet winner does not always exist, it is natural to require a voting rule to select the Condorcet winner whenever one exists; such voting rules are called *Condorcet-consistent*. it would be tempting to conjecture that sequential voting is Condorcet-consistent. However, Example 5 illustrates that this is not the case. Indeed, in this example $C$ is the Condorcet winner, yet the outcome is $\{A, B\}$.

Another counterintuitive phenomenon that distinguishes sequential voting with three or more candidates both from simultaneous voting and from sequential voting with two candidates is that a voter may vote for an alternative that ends up not winning.

**Example 6.** Consider an election with four voters $ACB$, $BCA$, $BCA$, $CAB$, voting in this order. Additionally, assume that voters 1, 2, and 4 are centrist voters, and voter 3 is a top-bottom voter. We claim that the equilibrium vote vector for this election is $(A, \bot, C, C)$ (and the resulting outcome is $C$). Indeed, it is easy to see that player 4 is playing his best response. Further, if player 3 abstains, the outcome is $\{A, C\}$, if he votes for $A$, the outcome is $\{A\}$, and if he votes for $B$, the outcome is also $\{A\}$ (this relies on voter 4 being a centrist voter). Thus, voter 3 is playing his best response as well.

Now, consider voter 2. If he votes for $A$, the outcome would necessarily contain $A$, and he prefers $C$ to any such outcome. If he votes for $B$, one can check that the best response to that by voter 3 is to vote for $B$, in which case $C$ votes for $A$ and the outcome is $\{A, B\}$ (note that by voting for $C$, voter 3 can enforce the outcome to be $C$, but this is not in his best interests). Since 2 is a centrist voter, this change is not beneficial for him. Finally, if 2 votes for $C$, the outcome is $\{C\}$, so this deviation lowers 2's payoff by $\varepsilon$.

It remains to consider player 1. If he abstains or votes for $B$, voters 2 and 3 can enforce the outcome to be $B$. Now, suppose that he votes for $C$. If players 2 and 3 vote for $B$, player 4 will respond by voting $C$, so the outcome is $\{B, C\}$. This is the second-best outcome for players 2 and 3, and they clearly cannot make $B$ the unique winner, so voting for $B$ is actually their best response in this case. Voter 1 prefers $\{C\}$ to $\{B, C\}$, so this deviation is not profitable for him.

Note that, interestingly, if the first voter switches to voting for $C$, this hurts $C$ rather than helps him. Intuitively, by voting for $A$, the first voter creates a threat to voters 2 and 3, which forces them to support the favorite choice of the last voter.

## 4.3 Computing the Equilibrium Vote Vector

In general, the problem of computing the vote vector that corresponds to a subgame-perfect Nash equilibrium of a given election appears to be computationally hard. Indeed, unlike in the simultaneous case, it is not even clear if this problem is in NP: given a vote vector, there is no obvious way to check that no player can profitably deviate from it: e.g., performing such a check for player $\pi(1)$ is effectively equivalent to computing the equilibrium vote vector in a game of size $n-1$.

In terms of providing an upper bound on the computational complexity of this problem, it is not hard to show that computing the equilibrium vote vector is in PSPACE. Indeed, this problem (or, more precisely, its generalization where we need to find the equilibrium vote vector given some fixed history) can be solved recursively, and increasing the recursion depth by one increases the space requirements by $O(m)$. However, we do not have a matching lower bound: in fact, we do not even have a proof that this problem is NP-hard (though we strongly suspect that this is the case).

On the positive side, just like in the simultaneous setting, it turns out that computing the equilibrium outcome is easy if the number of voters or the number of candidates is bounded by a constant.

**Theorem 6.** *We can compute an SPNE of a given sequential election $\pi$ with the set of voters $V$, $|V| = n$, the set of candidates $C$, $|C| = m$, and a collection of utility vectors $\mathbf{u}^i$, $i = 1, \ldots, n$ in time $O((m+1)^n m \log u_{\max})$, where $u_{\max} = \max\{u_j^i \mid i \in V, j = 1, \ldots, m\}$.*

*Proof.* We can design an algorithm with the desired running time by combining dynamic programming with backwards induction. Assume without loss of generality that $\pi(i) = i$ for all $i \in V$. For any $i = 1, \ldots, n$ and any vote vector $\mathbf{r}^{i-1} = (r_1, \ldots, r_{i-1}) \in B^{i-1}$ (where we assume $B^0 = \emptyset$), let $A(\mathbf{r}^{i-1}, i)$ and $W(\mathbf{r}^{i-1}, i)$ denote, respectively, the SPNE strategy of the voter $i$ given a history $\mathbf{r}^{i-1}$ and the eventual election winner given that the first $i-1$ voters vote according to $\mathbf{r}^{i-1}$. For each voter $i$, there are exactly $(m+1)^{i-1}$ possible histories, so altogether there are $\frac{(m+1)^n - 1}{m}$ histories to be considered. Clearly, the values $A(\mathbf{r}^{i-1}, i)$, $i = 1, \ldots, n$, completely describe the players' equilibrium strategies, and the final outcome of the election is given by $W(\emptyset, 1)$. It remains to describe how to compute the quantities $A(\mathbf{r}^{i-1}, i)$ and $W(\mathbf{r}^{i-1}, i)$. This can be done by backwards induction, as follows.

Clearly, $A(\mathbf{r}^{n-1}, n)$ and $W(\mathbf{r}^{n-1}, n)$ are easy to compute for all $\mathbf{r}^{n-1} \in B^{n-1}$. Indeed, for any given history $\mathbf{r}^{n-1}$, player $n$ can simply compare the outcomes of all possible votes (including $b_n = \bot$), let $A(\mathbf{r}^{n-1}, n)$ be the vote that results in the best of those outcomes for player $n$, and let $W(\mathbf{r}^{n-1}, n)$ be the outcome itself. Now, suppose we have computed $A(\mathbf{r}^i, i+1)$ and $W(\mathbf{r}^i, i+1)$ for all $\mathbf{r}^i \in B^i$ and some $i < n$. To compute $A(\mathbf{r}^{i-1}, i)$ for some fixed history $(r_1, \ldots, r_{i-1}) \in B^{i-1}$, player $i$ needs to consider the reaction of player $i+1$ to each of his possible actions. That is, if $i$ were to submit a ballot $b_i$, the final outcome would be given by $W((r_1, \ldots, r_{i-1}, b_i), i+1)$. Thus we can set $A(\mathbf{r}^{i-1}, i)$ to the action that leads to the best possible outcome from player $i$'s perspective, and let $W(\mathbf{r}^{i-1}, i)$ to be this outcome.

Thus, each $A(\mathbf{r}^{i-1}, i)$ and $W(\mathbf{r}^{i-1}, i)$, $i = 1, \ldots, n$, $\mathbf{r}^{i-1} \in B^i$ can be computed by comparing $m+1$ possible outcomes, i.e., in time $O(m^2 \log u_{\max})$, so the theorem follows. □

Our second algorithm makes use of the fact that the equilibrium strategy of player $i$ only depends on the number of votes for each candidate submitted in rounds $1, \ldots, i-1$.

**Theorem 7.** *We can compute an SPNE of a given sequential election $\pi$ with the set of voters $V$, $|V| = n$, the set of candidates $C$, $|C| = m$, and a collection of utility vectors $\mathbf{u}^i$, $i = 1, \ldots, n$ in time $O(n^{m+1} m^2 \log u_{\max})$, where $u_{\max} = \max\{u_j^i \mid i \in V, j = 1, \ldots, m\}$.*

*Proof.* Again, we can assume without loss of generality that $\pi(i) = i$ for all $i \in V$. Let $A(n_1, \ldots, n_m, i)$ and $W(n_1, \ldots, n_m, i)$ denote, respectively, the SPNE vote of player $i$ and the eventual outcome of the election, given that in rounds $1, \ldots, i-1$ there were $n_j$ votes submitted for candidate $c_j$, $j = 1, \ldots, m$.

Clearly, it suffices to compute the values of $A(n_1, \ldots, n_m, i)$ and $W(n_1, \ldots, n_m, i)$ for $i = 1, \ldots, n$ and $n_j = 0, \ldots, n-1$ for $j = 1, \ldots, m$, i.e., $2n \times n^m$ quantities. Given these values, we can easily compute the actual equilibrium vote of each player, and the final outcome is given by $W(0, \ldots, 0, 1)$. Finally, the quantities $A(n_1, \ldots, n_m, i)$ and $W(n_1, \ldots, n_m, i)$ for $i = 1, \ldots, n$ and $(n_1, \ldots, n_m) \in \{0, \ldots, n-1\}^m$ can be computed similarly to the quantities $A(\mathbf{r}^{i-1}, i)$ and $W(\mathbf{r}^{i-1}, i)$ in the proof of Theorem 6. Each such computation can be performed in time $O(m^2 \log u_{\max})$, so the theorem follows. □

Note that one has to be careful when stating the computational problem in the sequential setting: we emphasize that our goal is to compute the election outcome (i.e., the set of winners) or the equilibrium vote vector, but not the equilibrium strategies themselves. Indeed, to specify the strategy of voter $\pi(n)$, we would have to list his responses to all possible histories, and the number of such histories is, in general, exponential in the input size. In particular, this means that, given a collection of strategies $(s_1, \ldots, s_n)$, one can check in time that is polynomial in the size of this collection whether $(s_1, \ldots, s_n)$ is an SPNE.

## 5 Simultaneous and Sequential Voting: A Comparison

In this section, we summarize the differences and similarities between simultaneous and sequential voting.

An important advantage of sequential voting is that it always has an equilibrium in pure strategies. Moreover, while this equilibrium is not necessarily unique, all equilibria of a given sequential voting game result in the same outcome. In contrast, a simultaneous voting game may not have a pure strategy Nash equilibrium (though it always has an equilibrium in mixed strategies). For example, as argued in Section 3, if the number of voters is a prime, and there is no candidate that is ranked first by all voters, Theorems 1 and 2 imply that the simultaneous voting game will not have a PNE. Similarly, if there are two alternatives, and one of them is preferred by a strict majority of voters, simultaneous voting does not have a PNE, whereas sequential voting is guaranteed to choose the "right" candidate. Indeed, a simple probabilistic argument shows that for a fixed number of candidates and uniformly random preferences, the probability that an election has a PNE goes to 0 as the number of voters increases. Further, as shown by Example 1, the simultaneous voting game can have several Nash equilibria with different winners.

Example 1 also illustrates another issue with simultaneous voting, namely, that it may fail to elect the candidate preferred by all voters when one exists. In contrast, it is not hard to see that in this case the sequential voting procedure will necessarily result in this candidate winning, with all voters except for the last one abstaining, and the last voter voting for this candidate. Indeed, this vote vector allows the first $n-1$ players to achieve their maximal possible payoff, and the last player obviously maximizes his payoff by voting for his favorite candidate.

On the other hand, equilibria of simultaneous voting enjoy several natural properties: all voters are sincere, no voter ever votes for the candidate he ranks last (this is a simple consequence of Theorem 2), and no voter ever votes for a candidate that is not part of the winning outcome. These properties are shared by sequential voting with two candidates. However, Examples 5 and 6 show that when the number of voters exceeds two, sequential voting does not have any of those properties.

From computational perspective, computing the equilibrium is easy in both scenarios if the number of candidates or the number of voters is small. However, in the general case, simultaneous voting seems to be more tractable, as we can easily check whether a given ballot corresponds to a pure Nash equilibrium. Ironically, though, the only hardness result we have is for simultaneous voting; for sequential voting, we do not have a formal proof of hardness though we strongly suspect that this problem is at least NP-hard (and possibly PSPACE-hard). We believe that the counter-intuitive examples in Section 4 support our intuition: they show that players cannot cut the search space by applying natural heuristics, such as, e.g., assuming that everyone votes sincerely.

## 6 Related Work

There is a number of papers that consider voting from game-theoretic perspective, and study the relationship between the outcomes of simultaneous and sequential voting.

In particular, our model is somewhat similar to that of Sloth [17], who also views voting as a game of complete information, and studies both sequential and simultaneous voting. There are, however, several important differences between our work and that of [17]. First, Sloth considers the case of two alternatives, and then extends the results to multistage voting, where alternatives are eliminated one by one. Thus, in her model, the voters never have to choose from more than two alternatives at a time. Second, we assume that the voters have strict preferences over all possible outcomes, while paper [17] allows for indifferences in individual preferences. Third, [17] employs a deterministic tie-breaking rule. Finally, we allow the voters to abstain, while paper [17] does not. The latter distinction is important, since in simultaneous voting without abstention, there are equilibria in which all voters prefer $A$, yet vote for $B$, simply because no single agent can change the outcome by deviating. This issue is, in fact, the main focus of [17], which shows that one can get rid of such undesirable equilibria by introducing a small amount of uncertainty into the model. Thus, our approach can be viewed as providing an alternative method to eliminate some of the "bad" equilibria, namely, by allowing abstentions. Paper [17] also establishes a relationship between outcomes of sequential and simultaneous voting; however, due to differences in the model this relationship is different from the one that exists in our setting.

Subsequent papers in this vein, such as, e.g., [8, 18, 5, 11], focus on incomplete and asymmetric information settings. In particular, in an influential paper [5], Dekel and Piccione consider elections with two alternatives in which each voter observes a signal that provides him with some information about the values of the alternatives for him. For this setting, they prove that essentially any equilibrium of simultaneous voting is an equilibrium of sequential voting. A number of subsequent papers, such as, e.g., [2, 4, 1] consider different variants of the model proposed in [5]; however, to the best of our knowledge, none of these papers studies the setting with three or more candidates. Of those papers, the closest to ours is that of Battaglini [2], which modifies the model of [5] by allowing abstentions and has the same model of costly voting (arbitrarily small, but strictly positive cost) as we do. Our work can be seen as complementing this line of research: while we do not allow asymmetric information, many of our results (in particular, those for simultaneous voting) apply to any number of candidates. Further, our results for sequential voting with three or more candidates indicate that it will be difficult to generalize the existing results for the asymmetric information model to settings with three or more candidates. Indeed, in our model, the SPNE of sequential voting with three or more candidates fail to inherit the nice properties of the SPNE for two voters, such as sincerity or Condorcet consistency.

Costly voting is also studied by Borgers [3]. However, his focus is on comparing the social welfare under voluntary participation and that under compulsory participation. In addition, he does not assume that the voting costs are necessarily small, and only considers the case of two candidates.

Finally, there is a number of papers in the recent computational choice literature that consider multidimensional outcome spaces and multistage voting procedures, in which at each stage the voters collectively decide on a particular attribute of the outcome (see, e.g., [19, 14, 20, 21]). While such scenarios are often referred to as "sequential voting", they are very different from the sequential voting model considered in this paper: indeed, in our setting, each voter votes once only, while in the multistage voting each voter votes in every stage.

## 7  Conclusions and Future Work

We have studied the properties of equilibrium outcomes of Plurality voting with abstentions, for both simultaneous and sequential voting. This work can be extended in several directions. The most obvious open question is the complexity of computing an equilibrium outcome in sequential voting. While we strongly believe that this problem is hard, so far we have not been able to prove any lower bounds on its complexity. We would also like to develop a better understanding of the properties of such equilibria. For instance, we have demonstrated that they may fail to elect the Condorcet winner, but it remains unclear if they can elect a *Condorcet loser*, i.e., a candidate that loses a pairwise election to any other candidate.

Further afield, it would be interesting to apply equilibrium analysis to other voting rules, such as Borda or Copeland, both in sequential and in simultaneous voting scenario. An intriguing question here is which of the properties that a voting rule is known to possess under truthful voting (such as, e.g., monotonicity or Condorcet-consistency) are preserved by considering an equilibrium outcome instead.

Finally, we would like to extend our analysis to other equilibrium concepts, such as, e.g., strong equilibrium. This would allow us to eliminate some unintuitive equilibria such as the second equilibrium in Example 1.